\documentclass{elsart}
\usepackage{graphicx}

\begin{document}
\begin{frontmatter}

\title{
Stochastic energy-cascade model for 1+1 dimensional fully developed 
turbulence
}

\author{
J\"urgen  Schmiegel$^{a}$,
Jochen Cleve$^{b}$,
Hans C.\ Eggers$^{c}$, 
}
\author{
Bruce R.\ Pearson$^{d}$, and
Martin Greiner$^{e}$
}

\address{$^{a}$Network for Mathematical Physics and Stochastics,  
               Aarhus University, DK--8000 Aarhus, Denmark;
               email: schmiegl@imf.au.dk }
\address{$^{b}$ICTP, Strada Costiera, 11, 34014 Trieste, Italy;
               email: cleve@ictp.trieste.it }
\address{$^{c}$Department of Physics, University of Stellenbosch,
               7600 Stellenbosch, South Africa;
               email: eggers@physics.sun.ac.za }
\address{$^{d}$School of Mechanical Materials, 
               Manufacturing Engineering and Management,  
               University of Nottingham, 
               Nottingham NG7 2RD, United Kingdom;
               email: bruce.pearson@nottingham.ac.uk }
\address{$^{e}$Corporate Technology,  
               Information{\&}Communications,     
               Siemens AG, D-81730 M\"unchen, Germany;
               email: martin.greiner@siemens.com }

\begin{abstract}
  Geometrical random multiplicative cascade processes are often used
  to model positive-valued multifractal fields such as the energy
  dissipation in fully developed turbulence.  We propose a dynamical 
  generalization describing the energy dissipation in terms of a
  continuous and homogeneous stochastic field in one space and one
  time dimension.  In the model, correlations originate in the overlap
  of the respective spacetime histories of field amplitudes. The 
  theoretical two- and three-point correlation functions are found to 
  be in good agreement with their equal-time counterparts extracted 
  from wind tunnel turbulent shear flow data.
\end{abstract}

\end{frontmatter}

\newpage 

Whenever strongly anomalous, intermittent fluctuations, long-range
correlations, multi-scale structuring and selfsimilarity go hand in
hand, the label `multifractality' is attached to the underlying
process.  While in this Letter we have in mind fully developed
turbulence of fluid mechanics \cite{FRI95}, such processes occur in
various other fields such as formation of cloud and rain fields in
geophysics \cite{SCH87}, internet traffic of communication network
engineering \cite{PAR00}, and stock returns in finance.
\cite{MUZ00}, to name but a few.  

Random multiplicative cascade models (RMCMs) are commonly used to
model and visualize such phenomena since they generally exhibit
multifractality and reproduce the abovementioned properties
\cite{FED88}. They are usually implemented through a scale-independent
cascade generator which produces a nested hierarchy of scales and
multiplicatively redistributes the local measure.

In fully developed turbulence, RMCMs have often been employed to model
the energy flux through inertial-range scales. Due to their
multiplicative nature, they can easily reproduce multifractal scaling
exponents associated with the energy dissipation \cite{MEN91}, the
latter representing the intermittency corrections \cite{FRI95}.
Although the link between such models and the Navier-Stokes equation
remains unclear, recent investigations on multiplier distributions
\cite{JOU99,JOU00} and scale correlations \cite{CLE00} have shown that 
RMCMs do appear to contain more truth than might reasonably be expected 
from their phenomenological basis.

Nevertheless, these discrete RMCMs are purely geometrical constructs
and incapable of describing causal dynamical effects of the turbulent
energy cascade. A generalization in this direction is clearly
desirable. Hence, in this Letter, we present a dynamical RMCM in
$1{+}1$ space-time dimensions which respects causality and
homogeneity. It is related to recent, related efforts 
\cite{SCH01,BAR02,MUZ02,SCH03}, but goes beyond them in its dynamical
interpretation. We first show how this model yields multifractal 
scaling for arbitrary $n$-point correlation functions, proceeding 
thereafter to compare equal-time two- and three-point correlation 
functions to their counterparts obtained from wind-tunnel turbulent 
shear flow data.

Our dynamical RMCM is constructed by analogy to the geometrical case,
in which the amplitude of the positive-valued energy-dissipation
field, resolved at the dissipation scale $\eta$, is defined as the
product of independently and identically distributed random weights
$q(l_j)$,
%-------------------------------------------------------------------
\begin{equation}
\label{one}
  \varepsilon(\eta)
    =  \prod_{j=1}^{J} q(l_j)
    =  \exp\left( \sum_{j=1}^{J} \ln{q(l_j)} \right) \,,
\end{equation}
%-------------------------------------------------------------------
where $l_j$ is an element of a nested hierarchy of scales $\eta=l_J
\leq l_j=L/\lambda^j \leq l_0=L$ with $0{\leq}j{\leq}J$ the ``cascade
generation'' and $\lambda{>}1$ the discrete scale step. The integral
length $L$ and the dissipation length $\eta$ represent, respectively,
the largest and smallest length scale of the process. The geometrical
RMCM furthermore requires $\langle q \rangle = 1$ because of 
conservation of energy flux.

We generalize (\ref{one}) by assuming that $\varepsilon$ is again
the multiplicative product of a stochastic field, but that this field
is now defined on continuous $1{+}1$ spacetime:
%-------------------------------------------------------------------
\begin{equation}
\label{two}
  \varepsilon(x,t)
    =  \exp\left\{ 
       \int_{-\infty}^{\infty}dt' \int_{-\infty}^{\infty}dx' 
       f(x-x',t-t') \gamma(x',t')
       \right\}
       \; ,
\end{equation}
%-------------------------------------------------------------------
where $f$ is the ``index function'' described below and by assumption
$\gamma(x,t) \sim S_\alpha((dxdt)^{\alpha^{-1}-1}\sigma,-1,\mu)$ is a
L\'evy-stable white-noise field with index $0{\leq}\alpha{\leq}2$
\cite{SAM94}. For $\alpha{=}2$, this corresponds to a non-centered
Gaussian white-noise field. From its characteristic function,
$\langle\exp\{n\gamma\}\rangle = \exp\{-(\sigma^\alpha n^\alpha)
/(\cos(\pi\alpha/2)) + \mu n\}$ with $\alpha{\neq}1$, the parameter 
$\mu$ is fixed to $\mu=\sigma^\alpha/\cos(\pi\alpha/2)$ in order to 
satisfy the requirement $\langle\exp\{\gamma\}\rangle=1$.

Causality, i.e.\ the requirement that $\varepsilon(x,t)$ depends on 
the past but not the future, dictates that the index function
$f(x-x',t-t')$ must be zero for $t-t'<0$. Demanding also spatial
symmetry around $x$, we are led to the form
%-------------------------------------------------------------------
\begin{equation}
\label{three} 
  f(x-x',t-t')
    =  \left\{ \begin{array}{ll} 
       1 & \qquad (\,0{\leq}t-t'{\leq}T, \; 
           -g(t-t'){\leq}x-x'{\leq}g(t-t') \,) \,, \\
       0 & \qquad (\mbox{otherwise}) \; .
       \end{array} \right. 
\end{equation}
%-------------------------------------------------------------------
As illustrated in Fig.~1, the causality cone $g(t-t')$ incorporates a
correlation time $T$ and a correlation length $L$ with $g(T)=L/2$. The
exponent of the ansatz (\ref{two}) can be thought of as a moving
average over the stable white-noise field.

According to (\ref{three}), the time integration in (\ref{two}) runs
over $0{\leq}t-t^\prime{\leq}T$. Since to any given time $t_j$ there
corresponds a length scale $l_j {=}2g(t_j)$, there is a joint
hierarchy of length and time scales, so that (\ref{two}) factorizes
into integrals of $\gamma$ over the separate slices shown in Fig.~1,
%-------------------------------------------------------------------
\begin{equation}
\label{four}
  q(l_j)
    =  \exp\left\{
       \int_{t-t_{j-1}}^{t-t_{j}}dt' 
       \int_{x-g(t-t')}^{x+g(t-t')}dx'\, \gamma(x',t')
       \right\}
       \; .
\end{equation}
%-------------------------------------------------------------------
In order to interpret $q(l_j)$ as a random multiplicative weight, its
probability density needs to be independent of scale. Since the
$\gamma(x',t')$ are i.i.d., the integration domain of (\ref{four})
must therefore be independent of the scale index $j$.  Together with
the the boundary conditions $g(T-\Delta{T}_L){=}L/2$ and
$g(\Delta{T}_\eta){=}\eta/2$, this fixes the causality cone to
%-------------------------------------------------------------------
\begin{equation}
\label{five}
  g(t-t')
    =  \frac{(L/2)}
            {1 + \frac{(L-\eta)}{\eta}
             \frac{(T-\Delta{T}_L-(t-t'))}
                  {(T-\Delta{T}_L-\Delta{T}_\eta)} }
       \; 
\end{equation}
%-------------------------------------------------------------------
for times $\Delta{T}_\eta{\leq}t-t^\prime{\leq}T-\Delta{T}_L$.  To
complete the picture, we need to specify $g(t-t')$ for $0 \leq t-t'
\leq \Delta{T}_\eta$ and $T-\Delta{T}_L \leq t-t' \leq T$. Since on
physical grounds we expect $\Delta{T}_\eta{\ll}T$, the simplest choice
is $\Delta{T}_\eta=0$. For the remaining parameter $\Delta{T}_L$, we
assume $\Delta{T}_L \ll T$; it will be specified more fully below.

The construction proposed in Eqs.\ (\ref{two})-(\ref{five}) guarantees
that the one-point statistics of the dynamical RMCM are identical to
its geometrical counterpart. In order to qualify for a complete
dynamical generalization, not only the one-point statistics but the
$n$-point statistics in general should match. Hence, we now consider
the equal-time two-point correlator with $t_1{=}t_2{=}t$ and
$\Delta{x}=x_2{-}x_1>0$,
%-------------------------------------------------------------------
\begin{eqnarray}
\label{six}
  R_{n_1,n_2}(\Delta{x})
    &=&  \frac{ \left\langle
                \varepsilon^{n_1}(x_1,t) \, \varepsilon^{n_2}(x_2,t)
                \right\rangle }
              { \left\langle \varepsilon^{n_1}(x_1,t) \right\rangle 
                \left\langle \varepsilon^{n_2}(x_2,t) \right\rangle}
         \nonumber \\
    &=&  \frac{ \left\langle D(\Delta{x})^{n_1+n_2} \right\rangle}
              { \left\langle D(\Delta{x})^{n_1} \right\rangle
                \left\langle D(\Delta{x})^{n_2} \right\rangle } \,.
\end{eqnarray}
%-------------------------------------------------------------------
As shown in Fig.~2a, the correlation between two points a distance
$\eta \le \Delta{x} \le L$ apart stems from the overlap region of the
two index functions $f(x'-x_1,t)$ and $f(x_2-x',t)$.  This explains
why, in the second step of (\ref{six}), $R$ can be written solely
in terms of integrals $D$ over the overlap region,
%-------------------------------------------------------------------
\begin{equation}
\label{seven}
  D(\Delta{x})
    =  \exp\left(
       \int_{t-T}^{t-g^{(-1)}(\Delta{x}/2)} dt'
       \int_{x_2-g(t-t')}^{x_1+g(t-t')} dx' \gamma(x',t')
       \right) \,,
\end{equation}
%-------------------------------------------------------------------
since the contributions from the non-overlapping regions are
statistically independent and hence factorize and cancel. Introducing
the spatio-temporal overlap volume
%-------------------------------------------------------------------
\begin{eqnarray}
\label{eight}
  V(\Delta{x})
    &=&  \int_{t-T}^{t-g^{(-1)}(\Delta{x}/2)} dt'
         \int_{x_2-g(t-t')}^{x_1+g(t-t')} dx' 
         \nonumber \\
    &=&  \frac{\eta{L}(T-\Delta{T}_\eta)}{(L-\eta)}
         \ln\left( \frac{L}{\Delta{x}} \right)
         - \left( 
         \frac{\eta(T-\Delta{T}_\eta)}{(L-\eta)} - \Delta{T}_L
         \right) (L-\Delta{x}) \,,
\end{eqnarray}
%-------------------------------------------------------------------
and employing basic properties of stable distributions \cite{SAM94}, 
the expectation of the $n$-th power of $D$ is found to be
%-------------------------------------------------------------------
\begin{equation}
\label{nine}
  \left\langle D(\Delta{x})^n \right\rangle
    = \exp\left( \frac{\sigma^\alpha}{\cos\frac{\pi\alpha}{2}} 
                 V(\Delta{x}) (n-n^\alpha) \right)
         \; .
\end{equation}
%-------------------------------------------------------------------
Defining the multifractal scaling exponents
$\tau(n)=\tau(2)(n{-}n^\alpha)/(2{-}2^\alpha)$, with
$\tau(2)=(\sigma^\alpha/\cos\frac{\pi\alpha}{2})(2{-}2^\alpha)
         \eta L(T{-}\Delta{T}_\eta)/(L{-}\eta)$, 
as well as $\tau[n_1,n_2]=\tau(n_1+n_2)-\tau(n_1)-\tau(n_2)$,
substitution of (\ref{nine}) into (\ref{six}) leads to the final 
expression for the equal-time two-point correlator:
%-------------------------------------------------------------------
\begin{eqnarray}
\label{ten}
  \lefteqn{
  R_{n_1,n_2}(\Delta{x}) =
  }
  \nonumber \\
    &=&  \left( \frac{L}{\Delta{x}} \right)^{\tau[n_1,n_2]}
         \exp\left[ -\tau[n_1,n_2]
         \left( 
         1 - \frac{(L-\eta)\Delta{T}_L}{\eta(T-\Delta{T}_\eta)} 
         \right)
         \left( 1-\frac{\Delta{x}}{L}\right) 
         \right]
         \; .
\end{eqnarray}
%-------------------------------------------------------------------
Equal-time two-point statistics of our dynamical RMCM in $1{+}1$
dimensions hence show multiscaling behavior for
$\eta{<}\Delta{x}{\ll}L$, in complete analogy to the findings of the
corresponding geometrical RMCM \cite{CLE03}. We also note that
setting $\Delta{T}_L = (T-\Delta{T}_\eta)\eta/(L-\eta)$ eliminates the
second factor in (\ref{ten}) leaving $R_{n_1,n_2}(\Delta{x}) =
(L/\Delta{x})^{\tau[n_1,n_2]}$ to scale rigorously.

Turning to temporal two-point correlations, we follow the same recipe
as for the above.  Correlations in this case arise from the overlap
volume illustrated in Fig.~2b, and lead, after an analogous
straightforward calculation, to
%-------------------------------------------------------------------
\begin{eqnarray}
\label{eleven}
  R_{n_1,n_2}(\Delta{t})
    &=&  \frac{ \left\langle
                \varepsilon^{n_1}(x,t_1) \varepsilon^{n_2}(x,t_2)
                \right\rangle }
              { \left\langle \varepsilon^{n_1}(x,t_1) \right\rangle 
                \left\langle \varepsilon^{n_2}(x,t_2) \right\rangle}
         \nonumber \\
    &=&  \left(
         \frac{\Delta{t}-\Delta{T}_L}{T-\Delta{T}_L}
         + \frac{\eta}{L} 
         \left(
         1 - \frac{\Delta{t}-\Delta{T}_L}{T-\Delta{T}_L} 
         \right)
         \right)^{-\tau[n_1,n_2](1-\Delta{T}_L/T)}
         \nonumber \\
    &\approx&  
         \left( 
         \frac{T}{\Delta{t}} 
         \right)^{\tau[n_1,n_2]}
\end{eqnarray}
%-------------------------------------------------------------------
with $\Delta{t}=t_2-t_1$ and $x_1{=}x_2{=}x$. For simplicity, the 
parameter $\Delta{T}_\eta$ has been set to zero. The last step of 
(\ref{eleven}), valid for $\Delta{T}_L{\ll}\Delta{t}{<}T$ and 
$\eta{\ll}L$ only, shows that the temporal two-point correlator has 
scaling exponents identical to those of the equal-time counterpart.

Although respective spatio-temporal overlap volumes are more
complicated, two-point spacetime correlations with both $\Delta{x}
\neq 0$ and $\Delta{t} \neq 0$ can also be derived; see Ref.\ 
\cite{SCH02} for a complete analysis. Here, we prefer to continue with
equal-time three-point correlations; their generalization to 
equal-time $n$-point correlations is straightforward and explicit
expressions can again be found in Ref.\ \cite{SCH02}. The corresponding 
overlap volumes, illustrated in Fig.~2c, represent the starting point 
for a calculation analogous to (\ref{seven})-(\ref{nine}), which leads 
to
%-------------------------------------------------------------------
\begin{eqnarray}
\label{twelve}
  \lefteqn{
  R_{n_1,n_2,n_3}(x_1,x_2,x_3) =
  }
  \nonumber \\
    &=&  \frac{ \left\langle
                \varepsilon^{n_1}(x_1,t) \varepsilon^{n_2}(x_2,t)
                \varepsilon^{n_3}(x_3,t)
                \right\rangle }
              { \left\langle \varepsilon^{n_1}(x_1,t) \right\rangle 
                \left\langle \varepsilon^{n_2}(x_2,t) \right\rangle 
                \left\langle \varepsilon^{n_3}(x_3,t) \right\rangle}
         \nonumber \\
    &=&  \left( 
         \frac{L}{x_3-x_1} 
         \right)^{\tau[n_1+n_2,n_3]-\tau[n_2,n_3]}
         \left( 
         \frac{L}{x_2-x_1} 
         \right)^{\tau[n_1,n_2]}
         \left( 
         \frac{L}{x_3-x_2} 
         \right)^{\tau[n_2,n_3]}
         \nonumber \\
    & &  \exp\left\{ -
         \left( 
         1 - \frac{(L-\eta)\Delta{T}_L}{\eta(T-\Delta{T}_\eta)}
         \right)
         \left[
         \left( \tau[n_1{+}n_2{,}n_3]{-}\tau[n_2{,}n_3] \right)
         \left( 1{-}\frac{x_3{-}x_1}{L} \right)
         \right. 
         \right. 
         \nonumber \\
    & &  \left.
         \left.
         \qquad
         + \tau[n_1{,}n_2] 
         \left( 1{-}\frac{x_2{-}x_1}{L} \right)
         + \tau[n_2{,}n_3] 
         \left( 1{-}\frac{x_3{-}x_2}{L} \right)
         \right]
         \right\}
\end{eqnarray}
%-------------------------------------------------------------------
with $x_1 < x_2 < x_3$ and $\eta \leq x_i-x_j \leq L$ for all
$i>j=1{,}2{,}3$. In the case of $n_1=n_2=n_3=1$ and small separations
$|x_i-x_j| \ll L$, or for all separations if the parameter
$\Delta{T}_L$ is fine-tuned, this simplifies to
%-------------------------------------------------------------------
\begin{eqnarray}
\label{thirteen}
  \lefteqn{
  R_{1,1,1}(x_2{-}x_1{=}\mbox{const}, x_3{=}x)
  } 
  \nonumber \\
    &\sim&  \left\{ \begin{array}{ll} 
            \left( \frac{L}{x-x_1} \right)^{\tau(2)}
            \left( \frac{L}{x_2-x} \right)^{\tau(2)}
            & \qquad (x_1 < x < x_2) \\
            \left( \frac{L}{x-x_1} \right)^{\tau(3)-2\tau(2)}
            \left( \frac{L}{x-x_2} \right)^{\tau(2)}
            & \qquad (x_1 < x_2 < x) \; .
            \end{array} \right. 
\end{eqnarray}
%-------------------------------------------------------------------
The comparative ease with which the three-point expressions
(\ref{twelve}) and (\ref{thirteen}) were derived can be traced back to
the fact that the present model incorporates spatio-temporal
homogeneity from the very beginning. This is to be contrasted with the
geometrical RMCMs of (\ref{one}) which, due to their hierarchical
structure, are not translationally invariant in space. This
non-invariance feeds through to all $n$-point observables and has to
be removed at considerable cost through successive spatial sampling
\cite{EGG01} before the latter can be compared to experimental
counterparts.

While we expect the model proposed in this Letter to find application
in many different phenomena, we demonstrate its qualities through
comparison with fully developed turbulence data. A velocity record of
the longitudinal component, obtained in a wind-tunnel shear flow
experiment \cite{PEA02}, was transformed into a one-dimensional
spatial record of the positive-valued surrogate energy dissipation
$\varepsilon(x)=15\nu(\partial_x v_x)^2$, where $\nu$ is the
viscosity. The sampled two-point correlations of order $n_1=n_2=1$ and
$n_1=2$, $n_2=1$ are plotted in Fig.~3.  Within the inertial range
$\eta \ll |x_2-x_1| \ll L$, the data reveals rigorous power-law
scaling with exponents $\tau[1,1]=0.184$ and $\tau[2,1]=0.34$. This
fixes the intermittency exponent $\tau(2)=0.184$ and the stable index
$\alpha=1.80$, which are the relevant model parameters for multifractal 
scaling. Once these have been fixed, no further room for adjustment is 
left for the theoretical three-point correlation (\ref{thirteen}), 
which is compared to its experimental counterpart in Fig.~4. 
Independent of the various combinations for the two-point distances 
$\eta \ll |x_i-x_j| \ll L$, the agreement between model and data is 
remarkable. This demonstrates that the proposed stochastic process, 
whose parameters have been fixed from lowest-order two-point 
correlations, is capable to describe the equal-time multivariate 
statistics of the turbulent energy dissipation beyond two-point order. 
It proofs also that the turbulent energy cascade can be thought of as 
a consistent multifractal process.

The dynamical RMCM presented here is a generalization of the
geometrical RMCM. By construction, it is causal, continuous and
homogeneous, does not make use of a discrete hierarchy of scales and
stochastically evolves a positive-valued field in one space and one
time dimension.  Several generalizations of this new model come to
mind immediately, such as the stochastic evolution in $n{+}1$ dimensions
with the optional inclusion of spatial anisotropy, the use of
independently scattered random measures to describe deviation from
log-stability \cite{BAR03}, the discretisation of space-time into 
smallest cells to model dissipation, and a dynamical RMCM for vector 
fields to model the turbulent velocity field.

%%%%%%%%%%%%%%%%%%%%%%%%%%%%%%%%%%%%%%%%%%%%%%%%%%%%%%%%%%%%%%%%%%%%%%%%%

\newpage
%-----------------------------------------------------------------------
%%%%%%%%%%%%%
% FIGURE 1  %
%%%%%%%%%%%%%
\begin{figure}
\includegraphics[width=\textwidth]{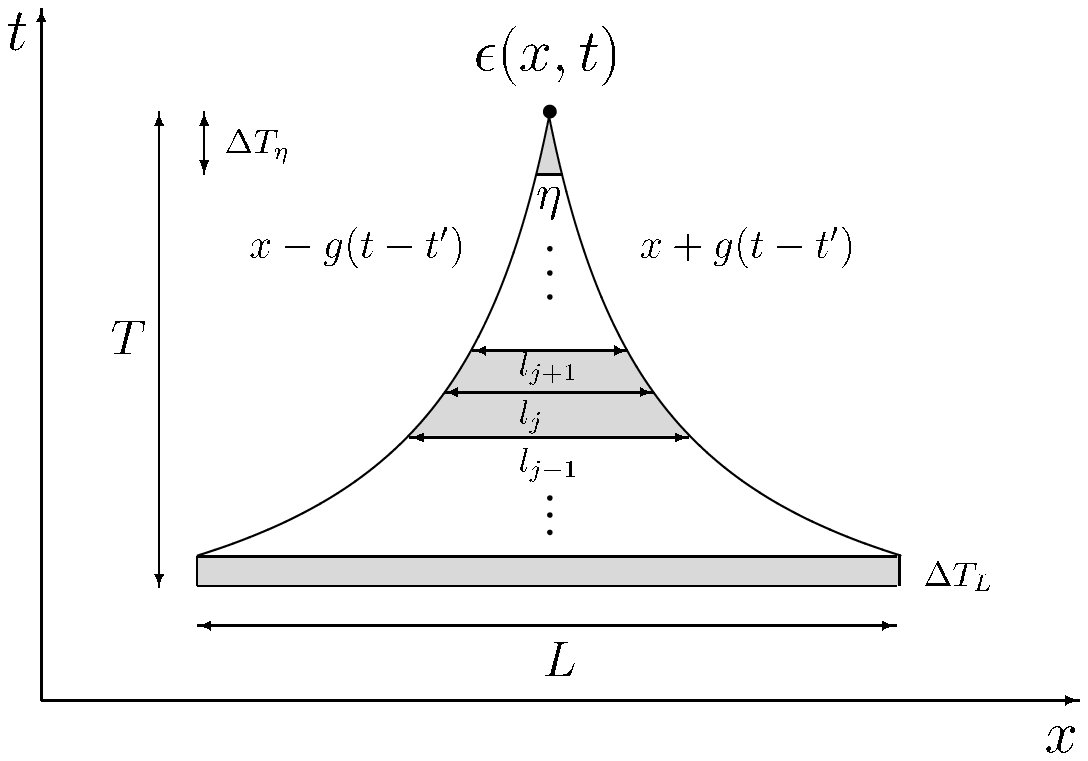}
\caption{
Causal space-time ``cone'' for the positive-valued
multifractal field $\varepsilon(x,t)$.  All field amplitudes
$\gamma(x',t')$ inside the causal space-time ``cone'' bordered by the
index function (\ref{three}) contribute multiplicatively to
$\varepsilon$. See text for further detail.
}
\end{figure}
%-----------------------------------------------------------------------

\newpage
%-----------------------------------------------------------------------
%%%%%%%%%%%%%
% FIGURE 2  %
%%%%%%%%%%%%%
\begin{figure}
\includegraphics[width=\textwidth]{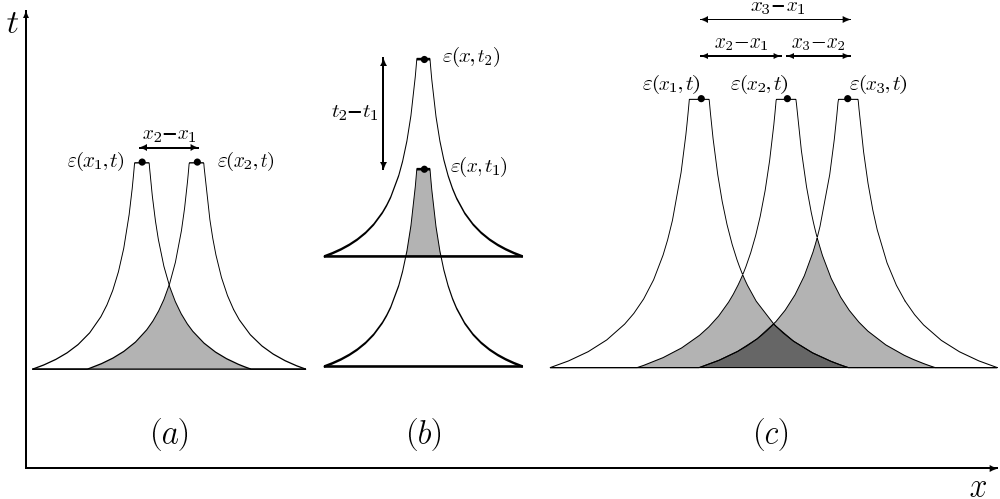}
\caption{
Spatio-temporal overlap volumes (shaded) producing the correlation for
the (a) equal-time and (b) temporal two-point correlator, as well as
(c) for the equal-time three-point correlator. To simplify
visualization, parameters have been set to $\Delta{T}_\eta=\Delta{T}_L=0$.
}
\end{figure}
%-----------------------------------------------------------------------

\newpage
%-----------------------------------------------------------------------
%%%%%%%%%%%%%
% FIGURE 3  %
%%%%%%%%%%%%%
\begin{figure}
\begin{center}
\includegraphics[width=10cm]{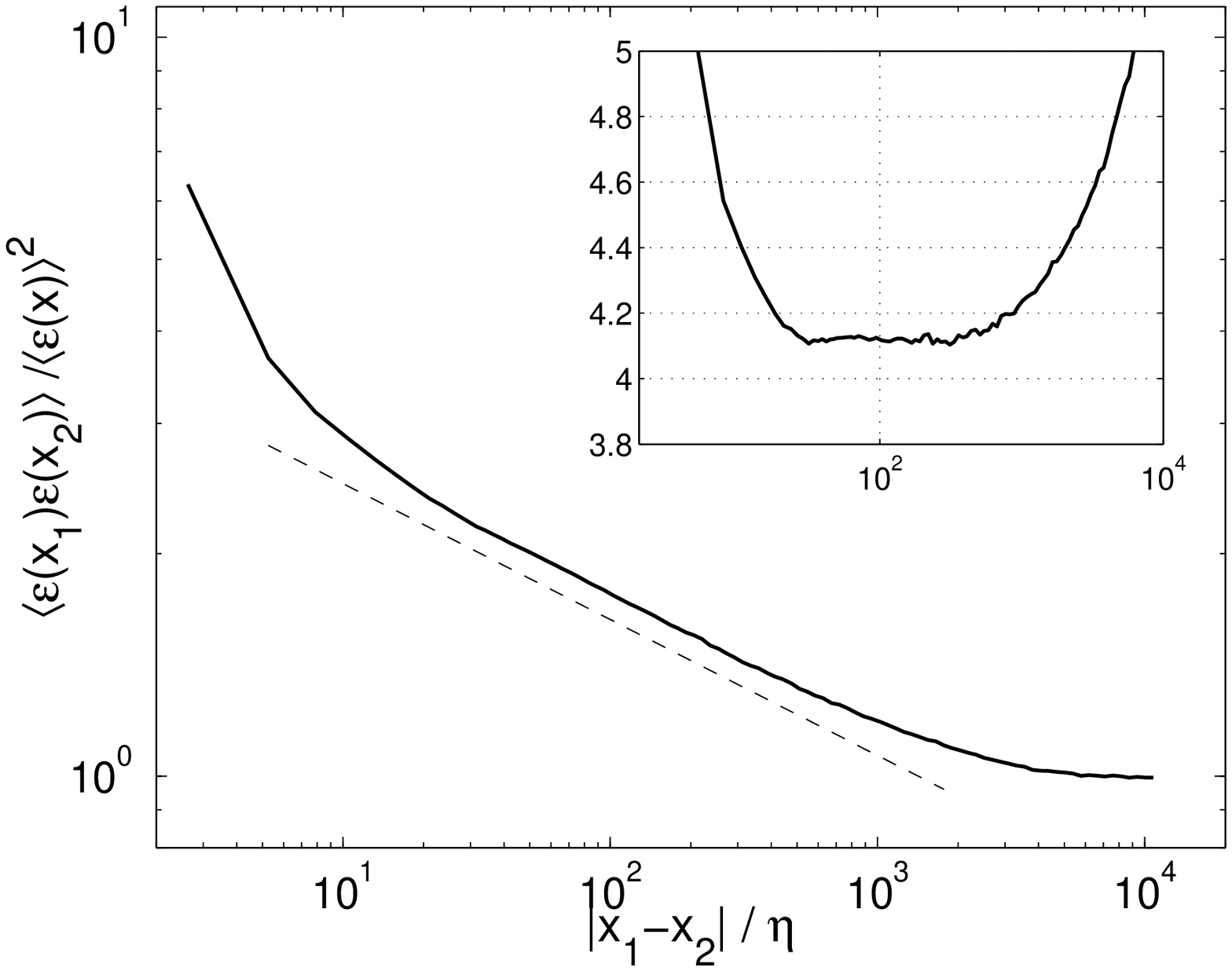}
\includegraphics[width=10cm]{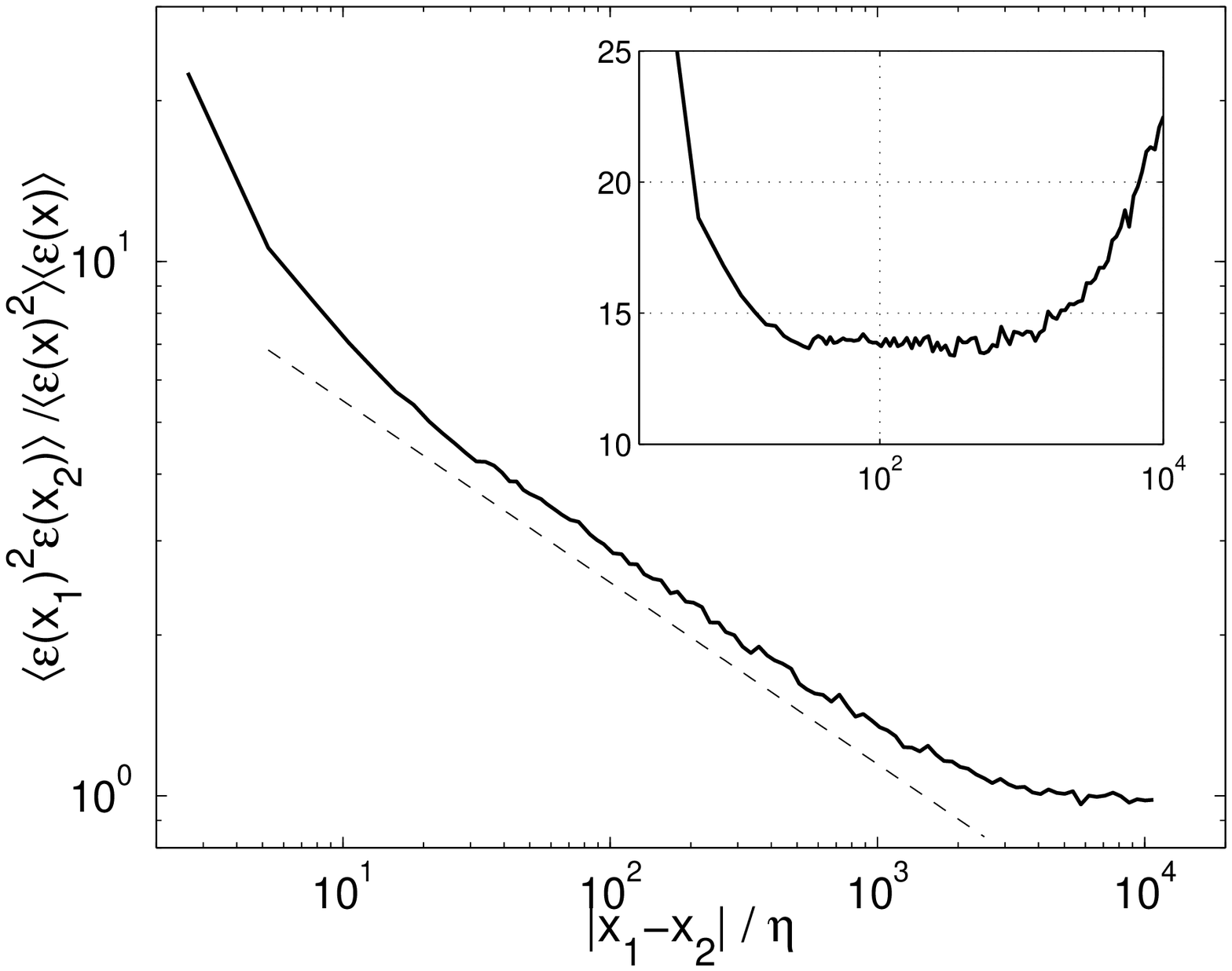}
\end{center}
\caption{
Two-point correlator $\langle \varepsilon^{n_1}(x_1,t)
\varepsilon^{n_2}(x_2,t) \rangle / ( \langle \varepsilon^{n_1}(x_1,t)
\rangle \langle \varepsilon^{n_2}(x_2,t) \rangle )$ of orders (a)
$n_1{=}n_2{=}1$ and (b) $n_1{=}2$, $n_2{=}1$ for the experimental
shear-flow dataset \cite{PEA02} with Taylor Reynolds number
$R_\lambda=860$ and integral length scale $L=1960\eta$ as a function
of the distance $|x_2-x_1|$ in units of the dissipation length $\eta$.
The insets represent the compensated plots, where the two-point
correlators have been divided by
$(|x_2-x_1|/\eta)^{-\tau[n_1,n_2]}$ with $\tau[1,1]=0.184$ and
$\tau[2,1]=0.34$, respectively.
}
\end{figure}
%-----------------------------------------------------------------------

\newpage
%-----------------------------------------------------------------------
%%%%%%%%%%%%%
% FIGURE 4  %
%%%%%%%%%%%%%
\begin{figure}
\begin{center}
\includegraphics[width=7cm]{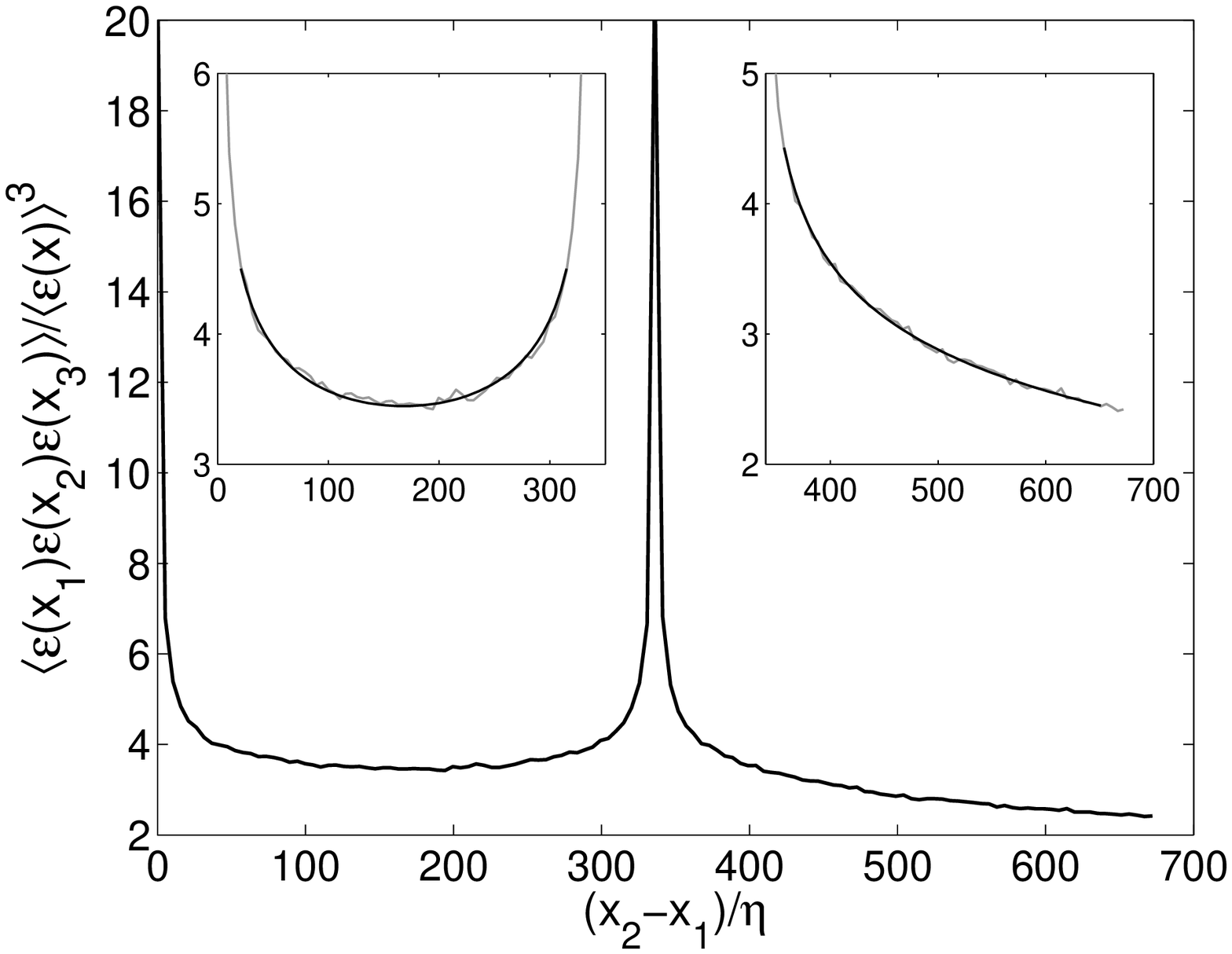} \\
\includegraphics[width=7cm]{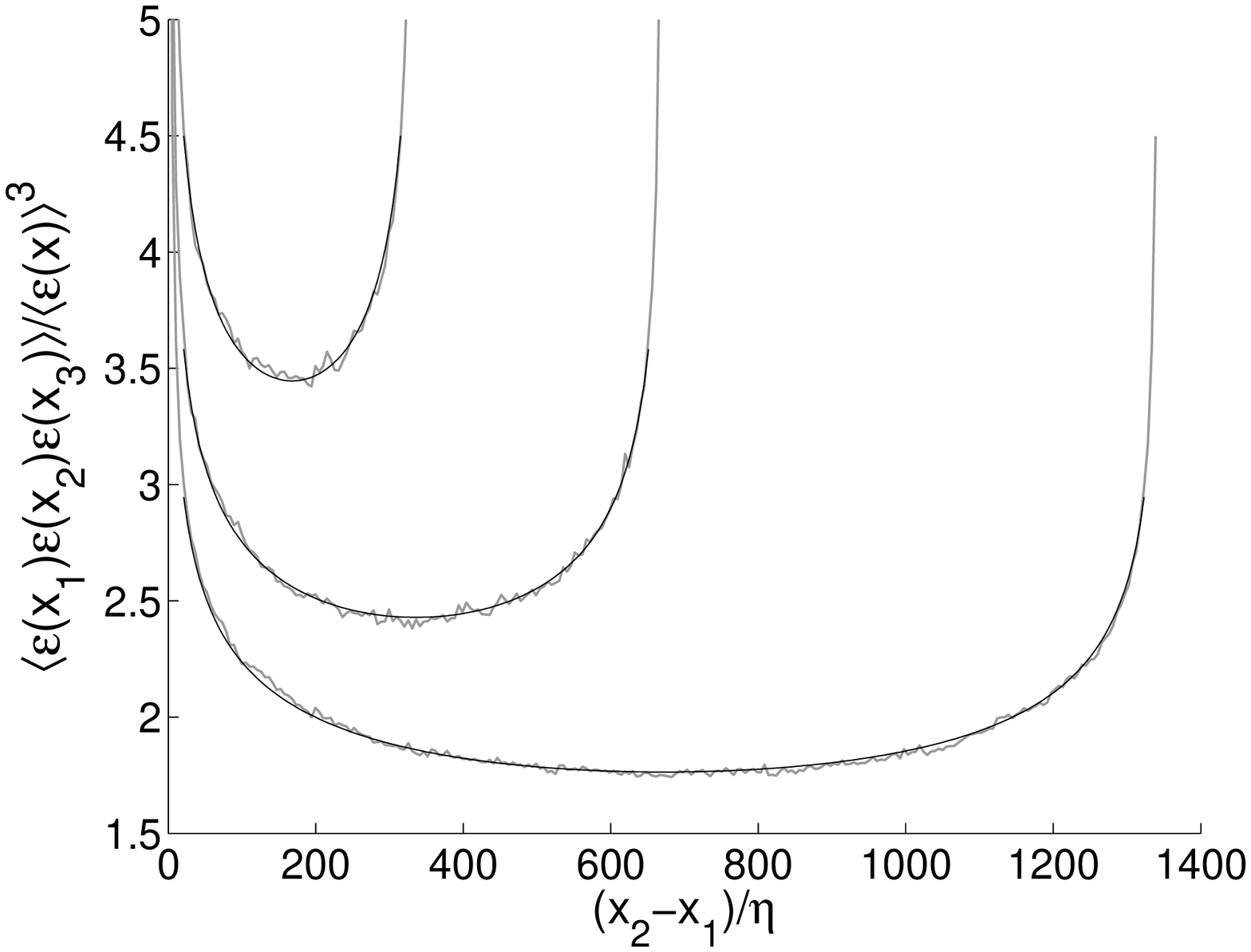} \\
\includegraphics[width=7cm]{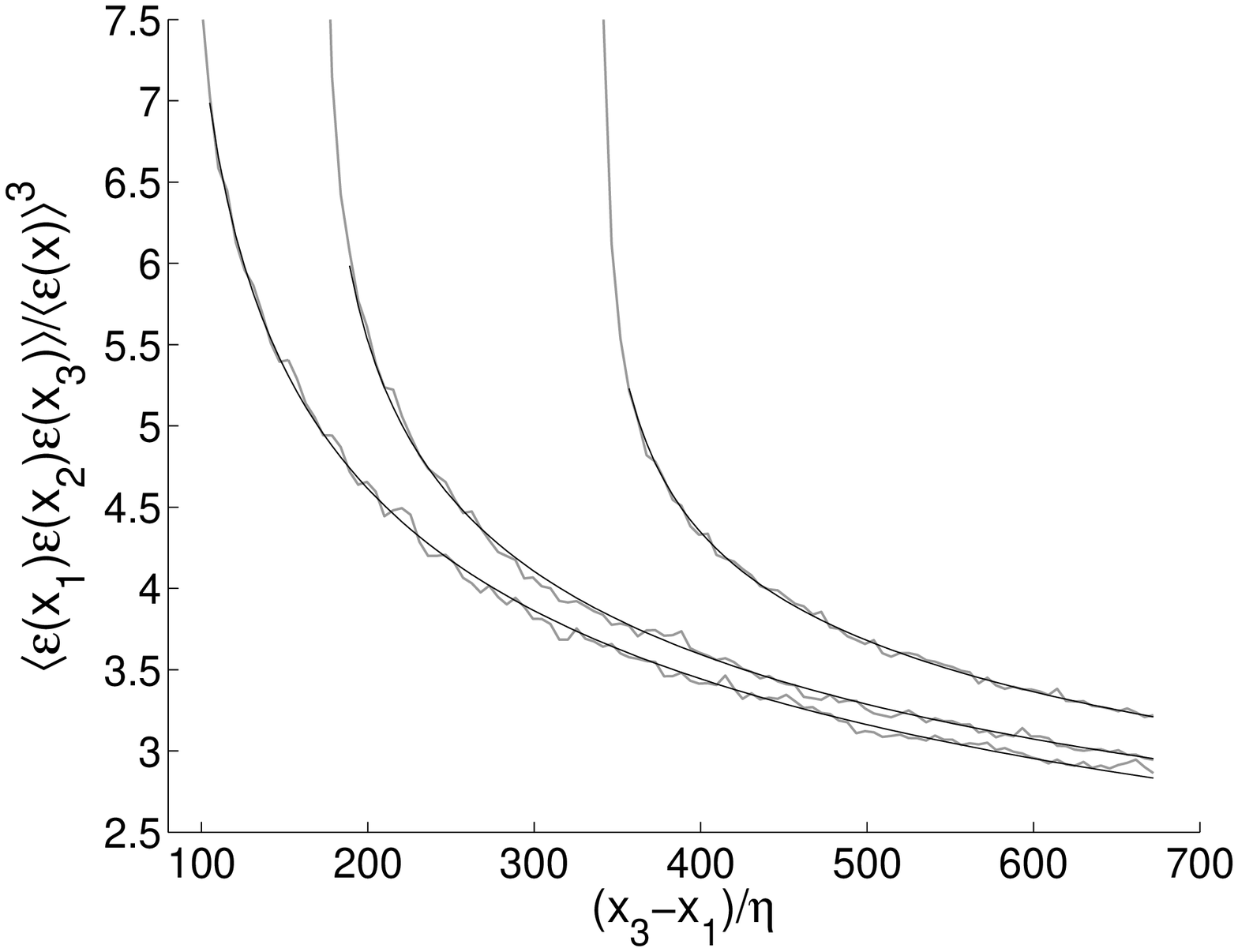}
\end{center}
\caption{
Comparison between expression (\ref{thirteen}) and the experimentally
extracted three-point correlator $\langle \varepsilon(x_1,t)
\varepsilon(x_2,t) \varepsilon(x_3,t) \rangle / \langle
\varepsilon(x,t) \rangle^3$. The same data set as in Fig.~3 has been
used. The scaling exponents $\tau[1,1]=0.184$ and $\tau[2,1]=0.34$
have already been fixed by the two-point correlators. In part (a), the
experimental three-point correlator is shown for a fixed two-point
distance $x_2{-}x_1=336\eta$. The two insets represent the comparison
with (\ref{thirteen}) for the two regimes $\eta < x_3{-}x_1 <
x_2{-}x_1$ and $x_3{-}x_1 > x_2{-}x_1$. Part (b) focuses on the regime
$\eta < x_2{-}x_1 < x_3{-}x_1$ with fixed $(x_3{-}x_1)/\eta=336$,
$672$ and $1344$, respectively, while (c) focuses on the regime
$x_3{-}x_1 > x_2{-}x_1$ with fixed $(x_2{-}x_1)/\eta=84$, $168$ and
$336$, respectively. For clarity, from left to right the curves of 
(c) have been shifted by $0$, $0.4$ and $0.8$ along the y-axis.
}
\end{figure}
%-----------------------------------------------------------------------

\end{document}